\documentclass[aps,prb,reprint,groupedaddress,showpacs]{revtex4-1}

\usepackage{graphicx}
\usepackage{dcolumn}
\usepackage{bm}
\usepackage{color}

\begin{document}

\preprint{APS/123-QED}

\title{Magnetic excitation emergence relevant to carrier scattering in semimetal Yb$_3$Ir$_4$Ge$_{13}$}	

\author{Kazuaki Iwasa}
\email{kazuaki.iwasa.ifrc@vc.ibaraki.ac.jp}
\affiliation{Frontier Research Center for Applied Atomic Sciences \& Institute of Quantum Beam Science, Ibaraki University, Tokai, Naka, Ibaraki 319-1106, Japan.}
\author{C.-L. Huang}
\author{Binod K. Rai}
\author{E. Morosan}
\affiliation{Department of Physics and Astronomy, Rice University, Houston, TX 77005, United States.}
\author{Seiko Ohira-Kawamura}
\author{Kenji Nakajima}
\affiliation{Materials and Life Science Division, J-PARC Center, Japan Atomic Energy Agency, Tokai, Naka, Ibaraki 319-1195, Japan}
\date{\today}

\begin{abstract}

Inelastic neutron scattering experiments have been conducted to investigate the semimetal system Yb$_3$Ir$_4$Ge$_{13}$. No clear crystal-field-split levels were observed up to an excitation energy of 60 meV, while a magnetic excitation appears at low energy below approximately 3~meV. This excitation shows a short spatial correlation, and an energy spectrum is reproduced by an imaginary part of generalized magnetic susceptibility on the basis of a damped harmonic oscillator model. The magnetic excitation emerges simultaneously with an anomalous electrical-resistivity enhancement below 20~K. The close relationship between the magnetic and transport behaviors evidences that the semimetal carriers are scattered by magnetic fluctuation dominated by the Yb $4f$ state.

\end{abstract}


\maketitle

\section{\label{sec:level1}Introduction\protect\\}

Heavy fermion (HF) behavior is a central issue in the research on strongly correlated electrons in rare earth-based materials~\cite{denseKondoEffect,HF_Grewe_Steglich}. In Ce-based intermetallics, interactions between $4f^1$ and conduction electrons give rise to various properties such as suppressed magnetic behaviors owing to a formation of Kondo singlet, semiconductor behaviors owing to a hybridization gap, valence fluctuation phenomena, etc. The $4f^{13}$ state of a Yb$^{3+}$ ion has also been investigated to be analogous to the Ce $4f^1$ state, corresponding to electron-hole symmetry. 

In this article, we discuss correlated electrons in the 3--4--13 class of materials $R_3T_4X_{13}$, where $R$ is a rare-earth element, $T$ stands for a transition-metal $d$-electron element, and $X$ is In, Ge, or Sn~\cite{Oswald_Yb3Rh4Sn13-type}. This class is of special interest because of its diverse physical properties, in particular their HF-like behavior~\cite{Ce3Co4Sn13_Cornelius,Ce3Co4Sn13_Slebarski_2012,Ce3Co4Sn13_LyleThomas,Ce3Co4Sn13_Slebarski_2013}.
Yb$_3$Ir$_4$Ge$_{13}$ crystallizes in a tetragonal structure, characterized by lattice constants of $a = b = 17.7674$ and $c=17.8229$~\AA~and the space group $I4_1/amd$ (No.~141)~\cite{Yb3RhCoIr4Ge13_PRB93,Yb3Ir4Ge13_fragilemagnetism}.
A magnetic susceptibility of Yb$_3$Ir$_4$Ge$_{13}$ shows an cusp-like anomaly at $T^\ast_{\rm mag} = 0.9$~K, which has been considered to be a signature of magnetic ordering. Based on the Curie-Weiss law analysis, a Weiss temperature was estimated to be $-17$~K, which indicates antiferromagnetic interactions between the Yb ions. An effective magnetic moment value was evaluated to be 4.2$\mu_{\rm B}$/Yb from the high-temperature data, which is close to the free-ion value of 4.52$\mu_{\rm B}$/Yb. 
At $T^\ast_{\rm mag}$, a specific heat scaled by temperature also exhibits a peak anomaly, which moves to higher temperature with magnetic fields applied parallel to the $c$ axis. The electrical resistivity increases on cooling in the whole measured temperature range (0.3--300~K). The data above 100~K is reproduced by an activation type behavior with an energy gap of 120~K, which had been considered to be attributed to a Kondo semiconductor or semimetal state. However, the resistivity deviates downward from this activation type, and shows a plateau-like behavior near 20~K, which indicates no distinct band gap formation. In addition, the resistivity increases rapidly below 20~K. 
Such unconventional behaviors remains unexplained, although it is readily apparent that the Kondo effect plays an important role in these electronic phenomena. Microscopic investigations of the electronic state, in particular the Yb $4f$-electron magnetic state, is highly required for further understanding. 

In the present study, we performed neutron spectroscopy in order to extract magnetic excitation in Yb$_3$Ir$_4$Ge$_{13}$ in the temperature range of the low temperature anomaly in the electrical resistivity. As a result, below approximately 20~K, we observed a drastic enhancement in the excitation below approximately 3~meV, which is closely relevant to the anomalies in the resistivity, the magnetic susceptibility, and the specific heat at $T^\ast_{\rm mag}$. 

\section{\label{sec:level2}Experimental details\protect\\}

Samples of $R_3$Ir$_4$Ge$_{13}$ ($R$: Yb and Lu) were synthesized as described in Ref.~5.
Several pieces of crystals (3.84 and 4.13~g of the Yb- and Lu-based materials, respectively) were selected for inelastic neutron scattering (INS) measurements. These were fixed on a aluminum plate using adhesive (CYTOP, the Asahi Glass Co., Ltd), the crystal-lattice direction of which were not coaligned. This plate was put inside an aluminum can filled with 1 atm of helium gas.

INS experiments were performed using the AMATERAS cold-neutron disk chopper spectrometer installed at the pulsed-neutron beam line BL14 at the Materials and Life Science Experimental Facility, J-PARC~\cite{AMATERAS_JPSJ}. Because of the almost identical sample mass of the Yb$_3$Ir$_4$Ge$_{13}$ and Lu$_3$Ir$_4$Ge$_{13}$ specimens, the Yb $4f$-electron contribution to the spectra was extracted from the difference between the data for these two compounds because of the non-magnetic Lu ions. A set of incident neutron energies were selected via chopper combination; $E_{\rm i}$ = 3.439, 5.246, 8.967, 18.69, and 60.39~meV. The energy resolutions corresponding to the full width at half maximum (FWHM) at the elastic scattering position were approximately 0.0633, 0.136, 0.272, 0.803, and 4.4~meV for these incident neutron energies, respectively. The Utsusemi software suite was used to analyze the spectral data obtained via the pulsed-neutron scattering technique~\cite{UTSUSEMI}.
Sample temperatures in the INS measurements were controlled between 0.6 and 50~K using a $^3$He closed-cycle refrigerator. 

\section{\label{sec:level3}Experimental results\protect\\}

\begin{figure}[b]
\includegraphics[width=4.2cm]{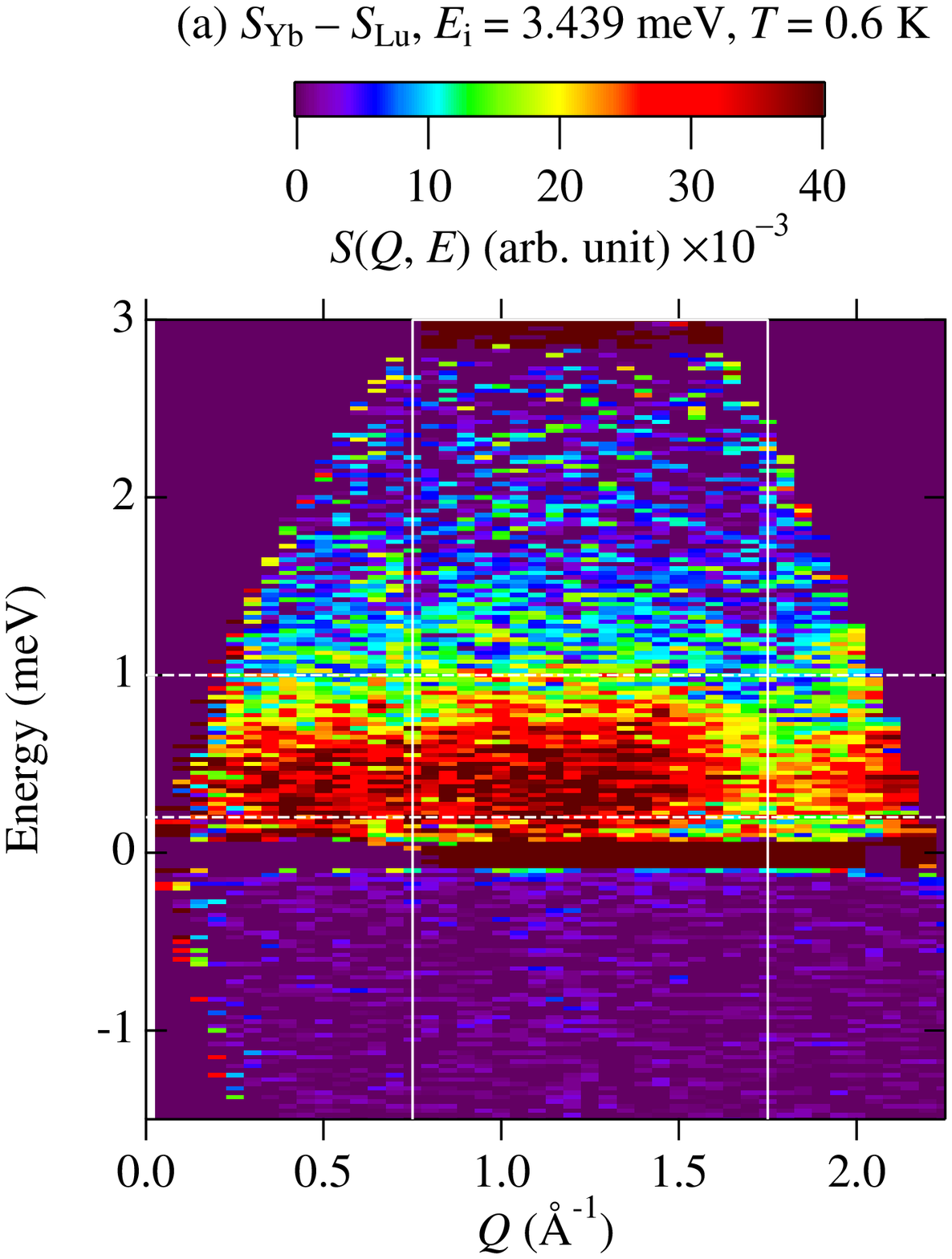}
\includegraphics[width=4.2cm]{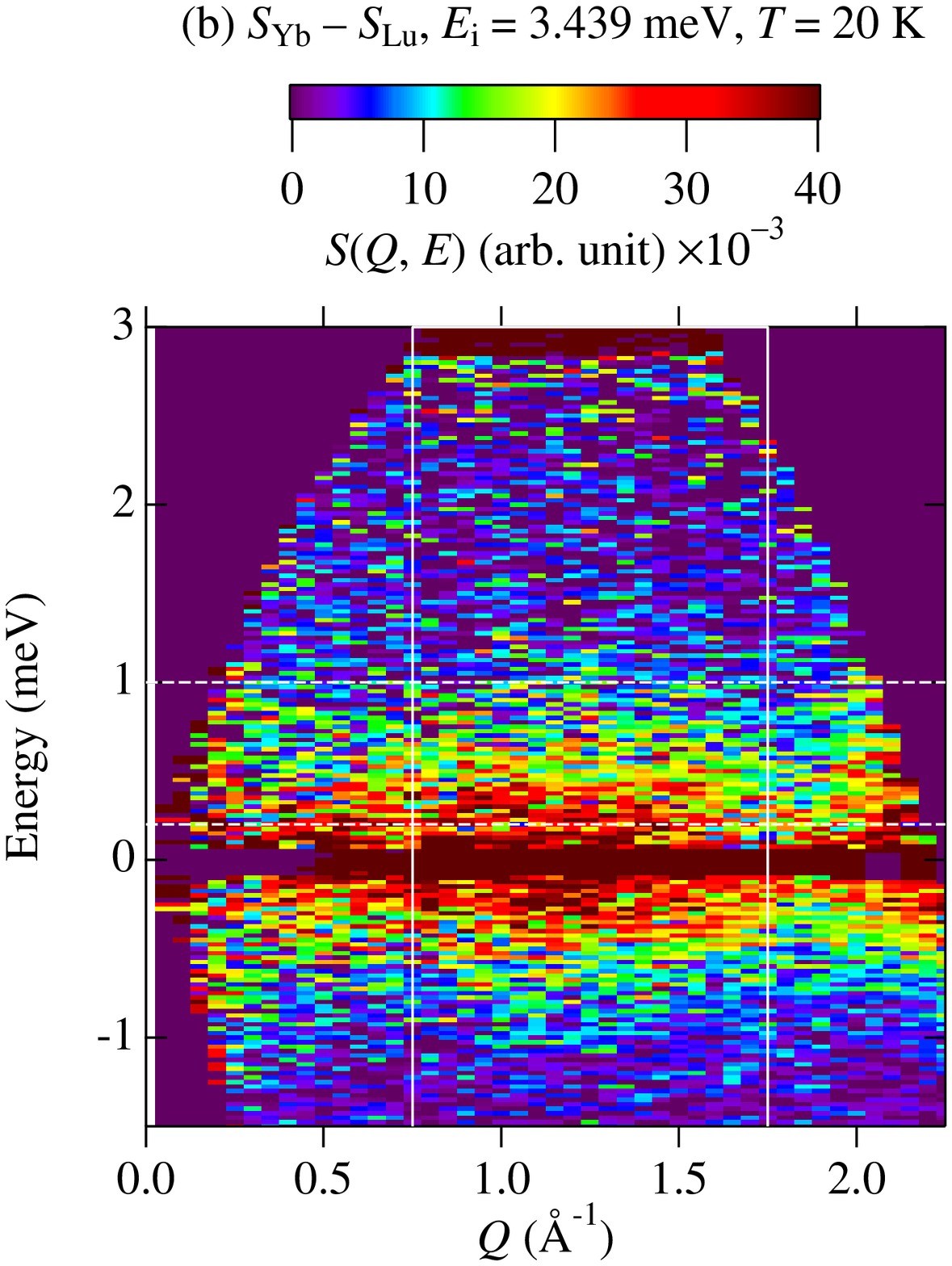}
\includegraphics[width=4.2cm]{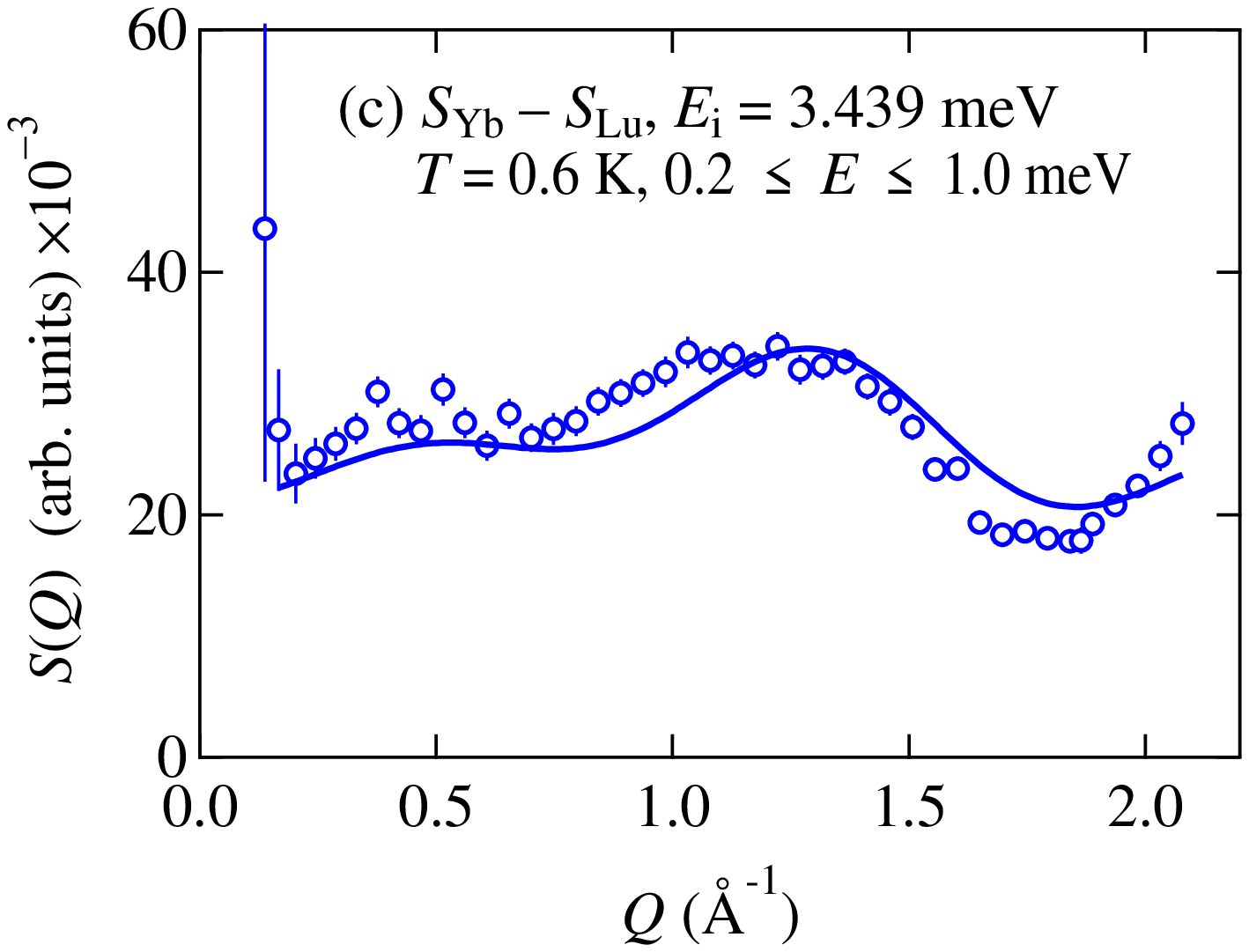}
\includegraphics[width=4.2cm]{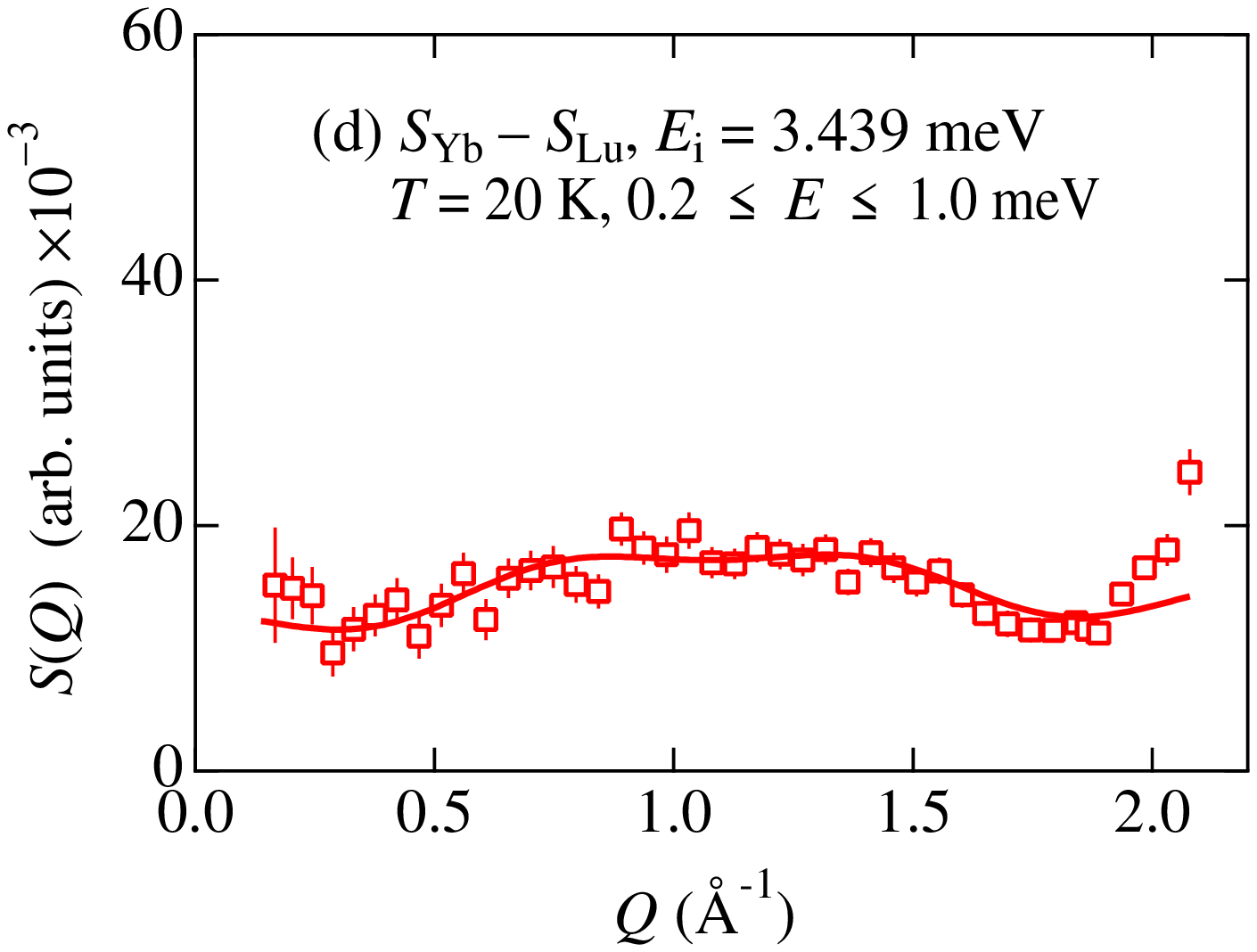}
\caption{\label{f_IQE} (Color online) Contour maps of neutron scattering function $S(Q,E)$ obtained from INS intensity differences between Yb$_3$Ir$_4$Ge$_{13}$ and Lu$_3$Ir$_4$Ge$_{13}$ at (a) 0.6~K and (b) 20~K. The incident neutron energy was $E_{\rm i}$ = 3.439 meV. 
The symbols in panels (c) and (d) show $S(Q)$ evaluated by the integrated intensity over the excited energy range between 0.2 and 1.0~meV [indicated by the vertical solid lines in (a) and (b)] at 0.6~K and at 20~K, respectively. The lines in (c) and (d) are the fitted results of the spin cluster model.
}
\end{figure}

Figure~\ref{f_IQE} shows contour maps of the neutron scattering function, $S(Q,E) \propto \left(\frac{k_{\rm f}}{k_{\rm }i}\right)^{-1}\frac{d^2\sigma}{d{\Omega}dE_{\rm f}}$, which were measured using $E_{\rm i}$ = 3.439 meV. The data were averaged with respect to the direction of scattering vector, ${\bm Q}$, because of the effectively polycrystal sample. 
$k_{\rm f}$ and $k_{\rm i}$ are outgoing and incident neutron wave vectors, respectively, and $\frac{d^2\sigma}{d{\Omega}dE_{\rm f}}$ is a second derivative of scattering cross section, $\sigma$. The result corresponds to the intensity differences between Yb$_3$Ir$_4$Ge$_{13}$ and Lu$_3$Ir$_4$Ge$_{13}$. 
Figure~\ref{f_IQE}(a) is the data taken at 0.6~K (below $T^\ast_{\rm mag}$ = 0.9 K), which shows strong INS signal around the excitation energy of $E=$ 0.4~meV. This signal is suppressed at 20~K (above $T^\ast_{\rm mag}$) and the intensities in the negative $E$ region becomes significant (anti-Stokes scattering), as seen in Fig.~\ref{f_IQE}(b). Such an enhancement of the intensities at lower temperature indicates magnetic excitations.
The INS signal of $S(Q)$ also depends on a scattering-vector magnitude, $Q$, as shown in Fig.~\ref{f_IQE}(c) and \ref{f_IQE}(d), which correspond to integrated intensity in the excited energy range between 0.2 and 1.0~meV. Stronger signal is seen around $Q=$ 1.2~\AA$^{-1}$, and the intensity becomes weaker near 0.75 and 1.75~\AA$^{-1}$. Magnitudes of the reciprocal lattice vectors of ${\bm Q}=$ (2, 0, 0), (2, 2, 0), and (2, 2, 2) are 0.707, 1.000, 1.224~\AA$^{-1}$. The $Q$ range of the observed INS signal covers these reciprocal lattice points, and thus the magnetic excitation is expected to extend over the Brillouin zone. Therefore, the magnetic excitation takes a shorter correlation length, which will be discussed later in detail. The decrease in $S(Q)$ with increasing temperature from 0.6 to 20~K indicates a suppression of the magnetic correlation.

\begin{figure}[b]
\includegraphics[height=7cm]{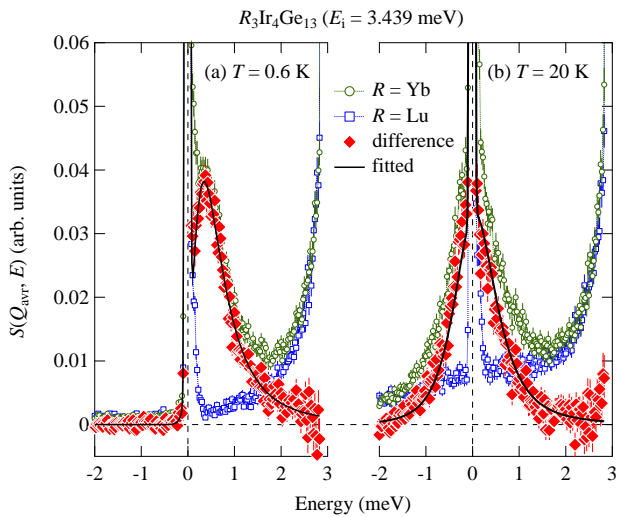}
\caption{\label{f_spectra} (Color online) Energy spectra at (a) 0.6 K and (b) 20 K. The data were taken using $E_{\rm i}=$ 3.439 meV and integration of the counts in the $Q$ range between 0.75 and 1.75~\AA$^{-1}$ [indicated by the horizontal broken lines in Figs.~\ref{f_IQE}(a) and \ref{f_IQE}(b)]. 
The green circles and blue squares are data of Yb$_3$Ir$_4$Ge$_{13}$ and Lu$_3$Ir$_4$Ge$_{13}$, respectively. The red diamonds correspond to the difference between these data, corresponding to magnetic signal form the Yb-based compound. The black solid lines indicate least-squares fitting results.}
\end{figure}

Symbols in Fig.~\ref{f_spectra} shows energy dependencies of $S(Q,E)$ at 0.6 and 20~K extracted from the data obtained using $E_{\rm i}=$ 3.439 meV for Yb$_3$Ir$_4$Ge$_{13}$ (green circles), Lu$_3$Ir$_4$Ge$_{13}$ (blue squares), and the difference $S_{\rm Yb}(Q,E) - S_{\rm Lu}(Q,E)$ (red diamonds).
These spectra were obtained by integrating the counts shown in Fig.~\ref{f_IQE} over $Q=0.75$ and 1.75~\AA$^{-1}$. The data for Lu$_3$Ir$_4$Ge$_{13}$ was used for estimating the background counts.
The intensity increases with increasing $E$ because of the tail of other $E_{\rm i}$. Almost pure magnetic scattering is obtained from Yb$_3$Ir$_4$Ge$_{13}$ after the subtraction of the Lu$_3$Ir$_4$Ge$_{13}$ data, as shown by the red filled diamonds. The spectrum at 0.6~K [panel \ref{f_spectra}(a)] shows a peak at 0.4~meV and a tail up to 3~meV. At 20~K [panel \ref{f_spectra}(b)], the peak is moved to lower energies, such that the maximum occurs close to $E = 0$, and the intensity in the negative $E$ region is enhanced, as seen in Fig.~\ref{f_IQE}(b). This feature will be analyzed by a neutron-scattering law in a discussion part.

\begin{figure}[t]
\includegraphics[height=5.1cm]{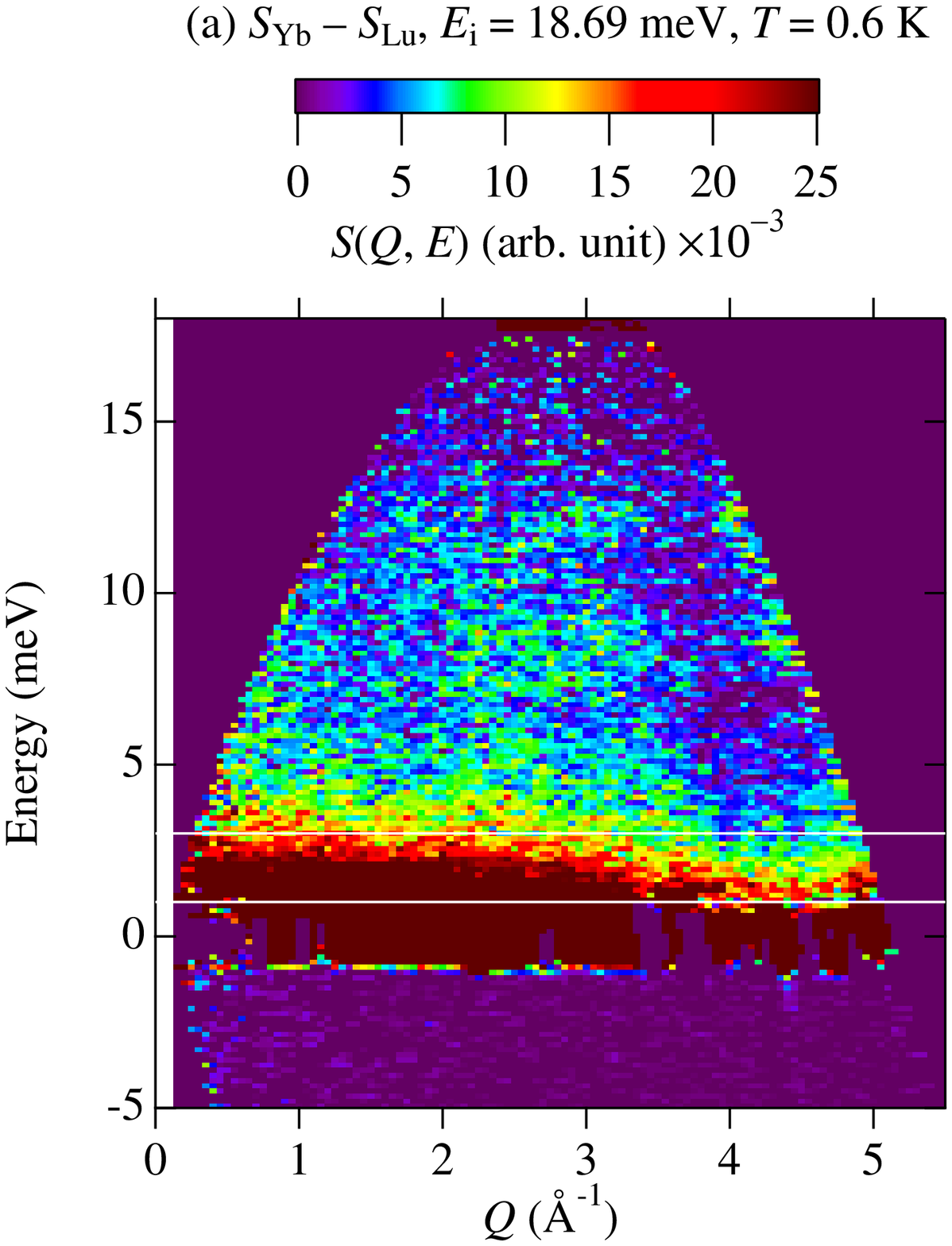}
\includegraphics[height=5.1cm]{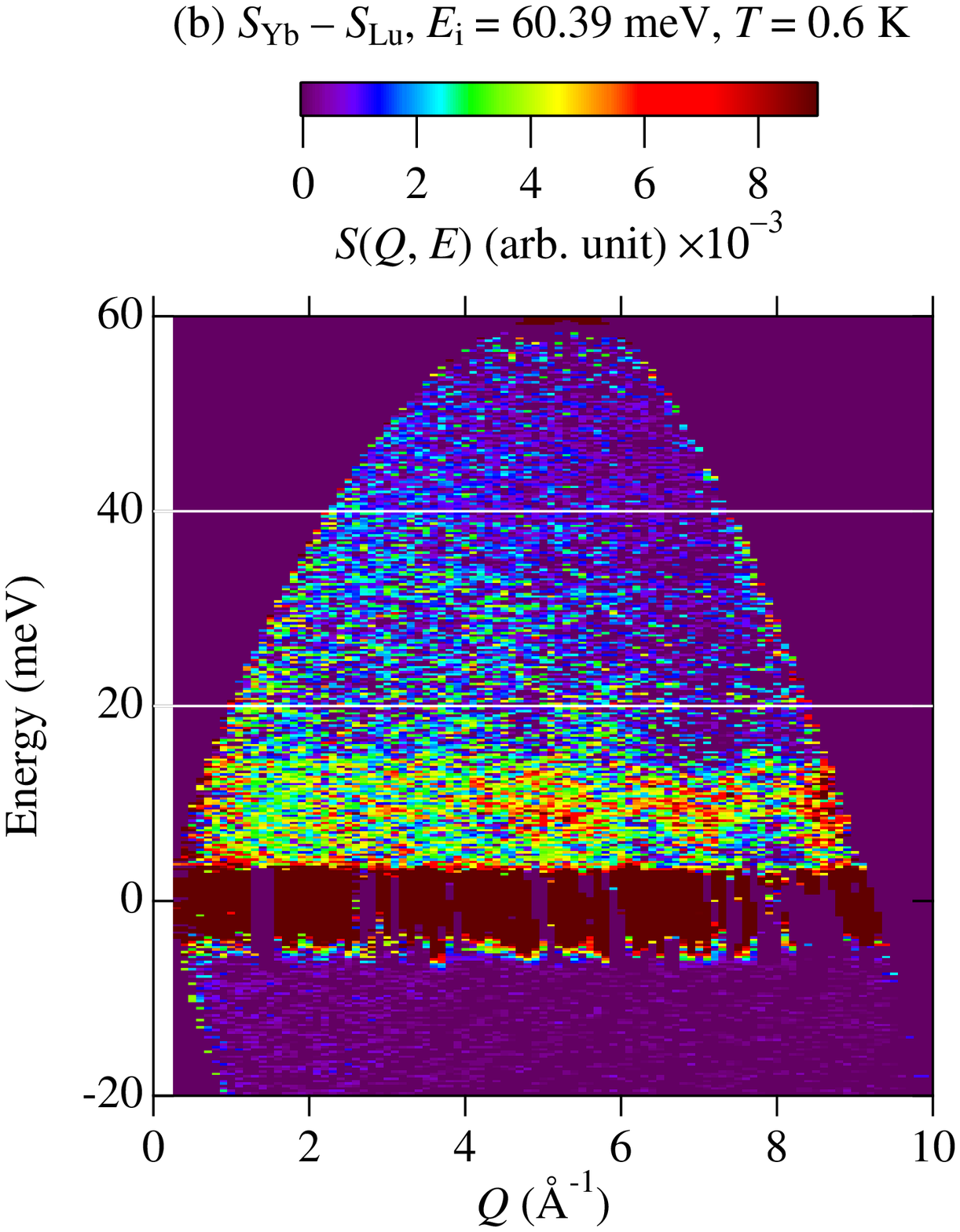}
\includegraphics[height=5.0cm]{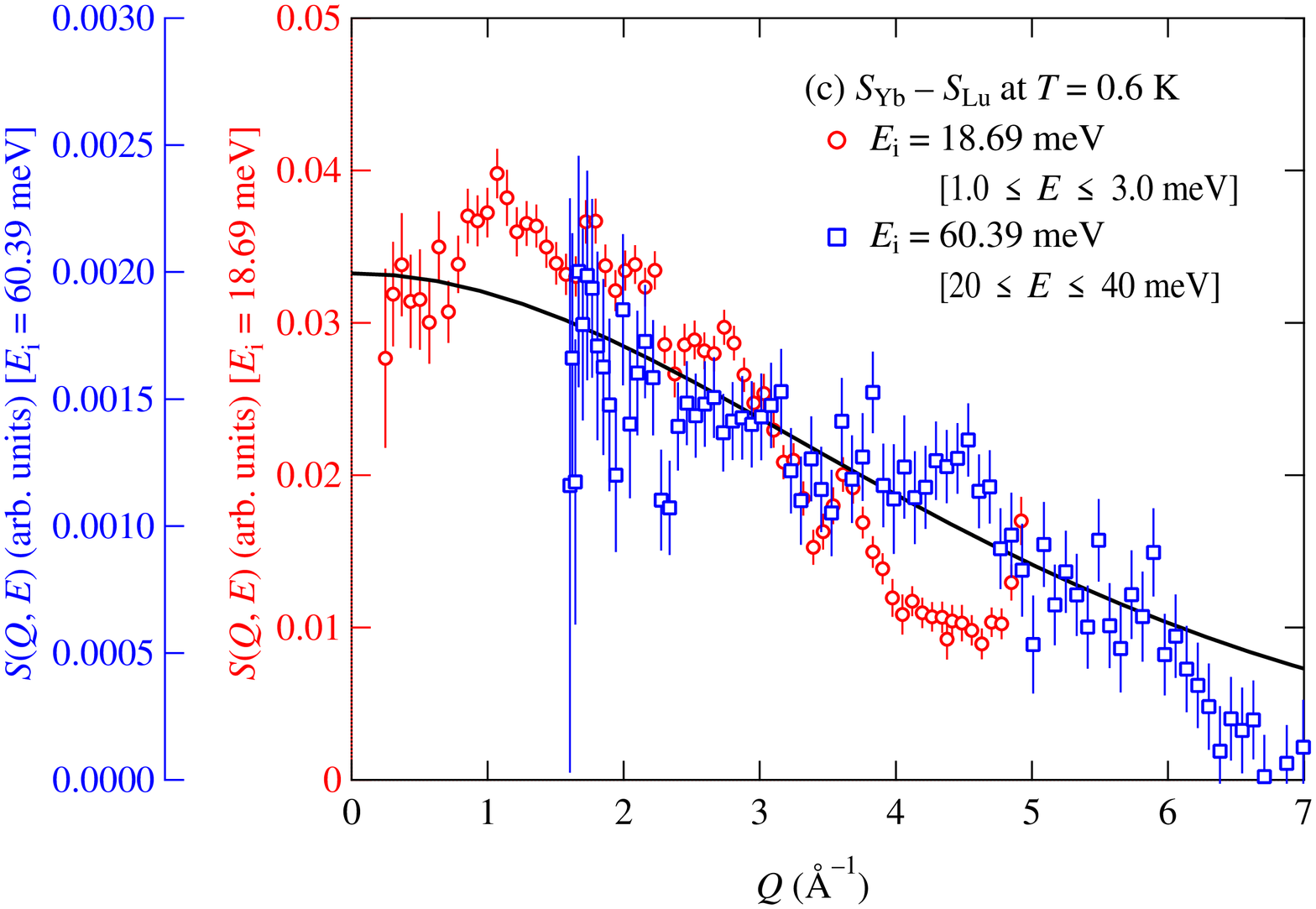}
\caption{\label{f_Qdep} (Color online) Contour maps of $S(Q,E)$ corresponding to the differences between Yb$_3$Ir$_4$Ge$_{13}$ and Lu$_3$Ir$_4$Ge$_{13}$ at 0.6~K using (a) $E_{\rm i}$ = 18.69 and (b) 60.39~meV, respectively.
(c) $Q$ dependencies of the intensity differences. The red circles are the intensities integrated within $E=$ 1.0--3.0~meV [indicated by the horizontal solid lines in (a)] of the dataset of $E_{\rm i}$ = 18.69 meV. The blue squares are those within $E=$ 20--40~meV [indicated by the horizontal solid lines in (b)] of the dataset of $E_{\rm i}$ = 60.39 meV. The black solid line is a squared magnetic form factor of a Yb$^{3+}$ $4f^{13}$ state~\cite{Pr_magff}.}
\end{figure}

The higher-$E_{\rm i}$ measurements were also conducted, in order to measure magnetic excitation in the higher excited-energy region. Figure~\ref{f_Qdep} shows the $S(Q,E)$ contour maps corresponding to the integrated-intensity differences between Yb$_3$Ir$_4$Ge$_{13}$ and Lu$_3$Ir$_4$Ge$_{13}$ at 0.6~K using $E_{\rm i}$ = 18.69 [panel (a)]  and 60.39~meV [panel \ref{f_Qdep}(b)], respectively.
The $E_{\rm i}$ = 18.69~meV data in Fig.~\ref{f_Qdep}(a) also show a strong intensity up to $E$ = 3~meV, which is consistent with the data using $E_{\rm i}$ = 3.439~meV shown in Fig.~\ref{f_IQE}(a). 
The red circles in Fig.~\ref{f_Qdep}(c) are integrated intensities in $E=$ 1.0--3.0~meV in the dataset of $E_{\rm i}$ = 18.69 meV, which decreases with increasing $Q$ while exhibiting rapid oscillations. The overall $Q$-dependence of $S(Q,E)$ is close to a squared magnetic form factor of a Yb$^{3+}$-ion $4f^{13}$ electron configuration~\cite{Pr_magff}, as shown by the black solid line in Fig.~\ref{f_Qdep}(c). This is indicative of a magnetic low-energy excitation below 3~meV.  

From the $E_{\rm i}$ = 60.39~meV data shown in Fig.~\ref{f_Qdep}(b), integrated intensity in $E=$ 20--40~meV was evaluated. The result is plotted by blue squares in Fig.~\ref{f_Qdep}(c). This high-energy INS signal also decreases with increasing $Q$, and is consistent with squared magnetic form factor of a Yb$^{3+}$ ion. This fact supports that broad magnetic scattering signal is located at the high-energy range. Another type of scattering intensity was detected up to 10~meV, which is more significant at higher $Q$ and shows a dispersion-like behavior. This signal can be attributed to phonon excitations. Since the coherent neutron scattering length of a Yb nucleus (${\bar b}=$ 12.43 fm) is larger than that of Lu (${\bar b}=$ 7.21 fm), Yb-ion motion might be apparent after taking the intensity difference between the two compound data.

\section{\label{sec:level4}Analysis\protect\\}

As mentioned above, a significant emergence of magnetic excitation below 3~meV has been detected by INS measurements. In order to extract parameters characterizing the dynamics, a least-squares fitting analysis was carried out for the data derived by integrating the counts in a range of $Q =$ 0.75--1.75~${\rm \AA}^{-1}$ obtained with $E_{\rm i}$ = 3.439~meV. As mentioned above, because the strong signal was observed in this $Q$ range covering the Brillouin zone, the integrated intensity allows us to discuss the dynamics with respect to the whole spatial correlation length. A general form of the neutron scattering law, $S(Q,E) \propto [n(E)+1] \chi''(Q,E)$, was taken into consideration, where $n(E)$ and $\chi''(Q,E)$ are the Bose--Einstein distribution function and the imaginary part of generalized magnetic susceptibility, respectively. We succeeded in reproducing the whole of observed spectra using a form
\begin{eqnarray}
\chi''(Q_{\rm avr},E) = \frac{\chi'_0 E_0^2 \gamma E}{(E^2-E_0^2)^2+(\gamma E)^2},
\label{eq1}
\end{eqnarray}
which corresponds to a damped harmonic oscillator model. Here, we denote an averaged wave vector, $Q_{\rm avr}$, because the data is an integration of intensity over the magnitude of ${\bm Q}$, as described above.

\begin{figure}[t]
\includegraphics[height=5.85cm]{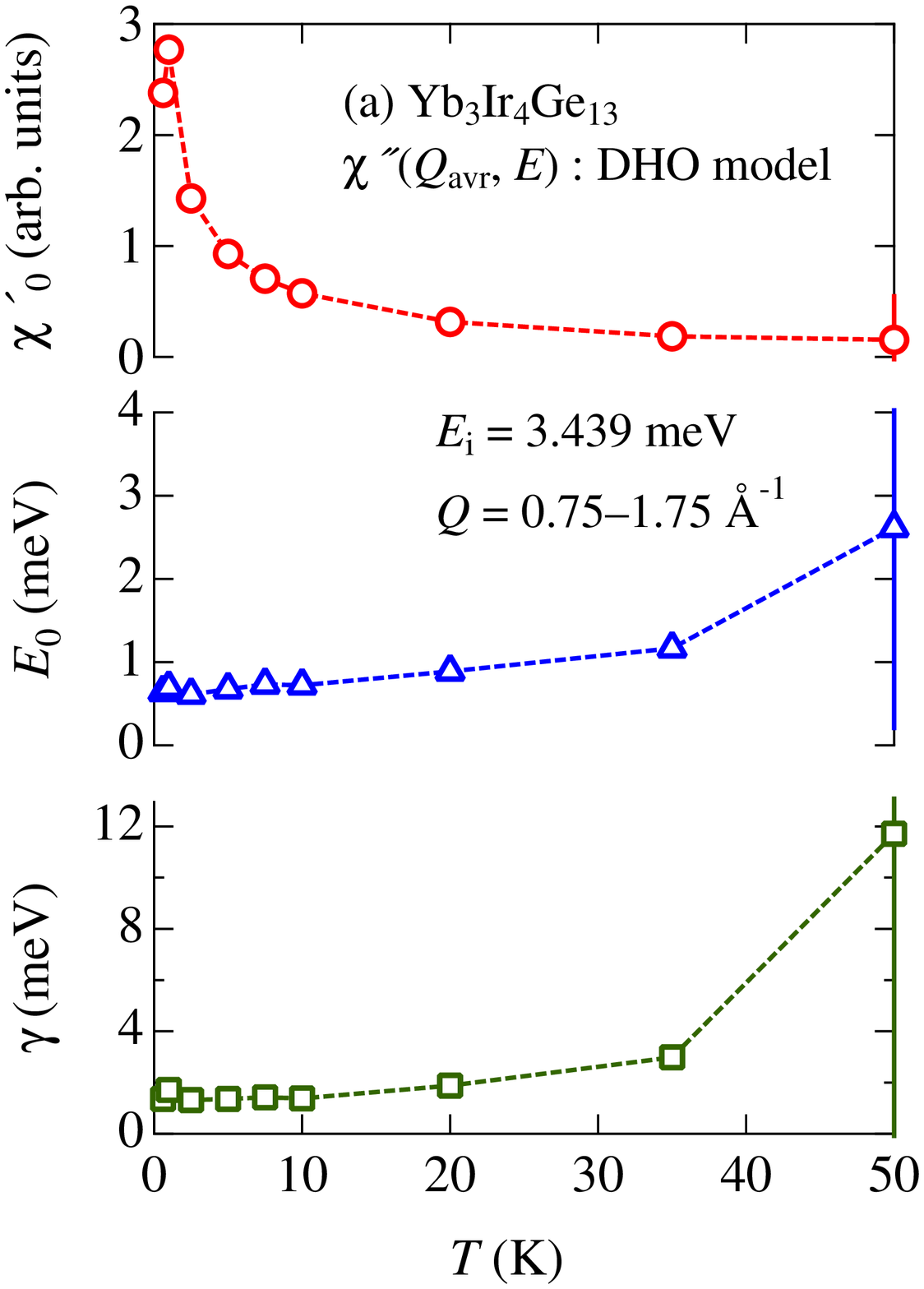}
\includegraphics[height=5.85cm]{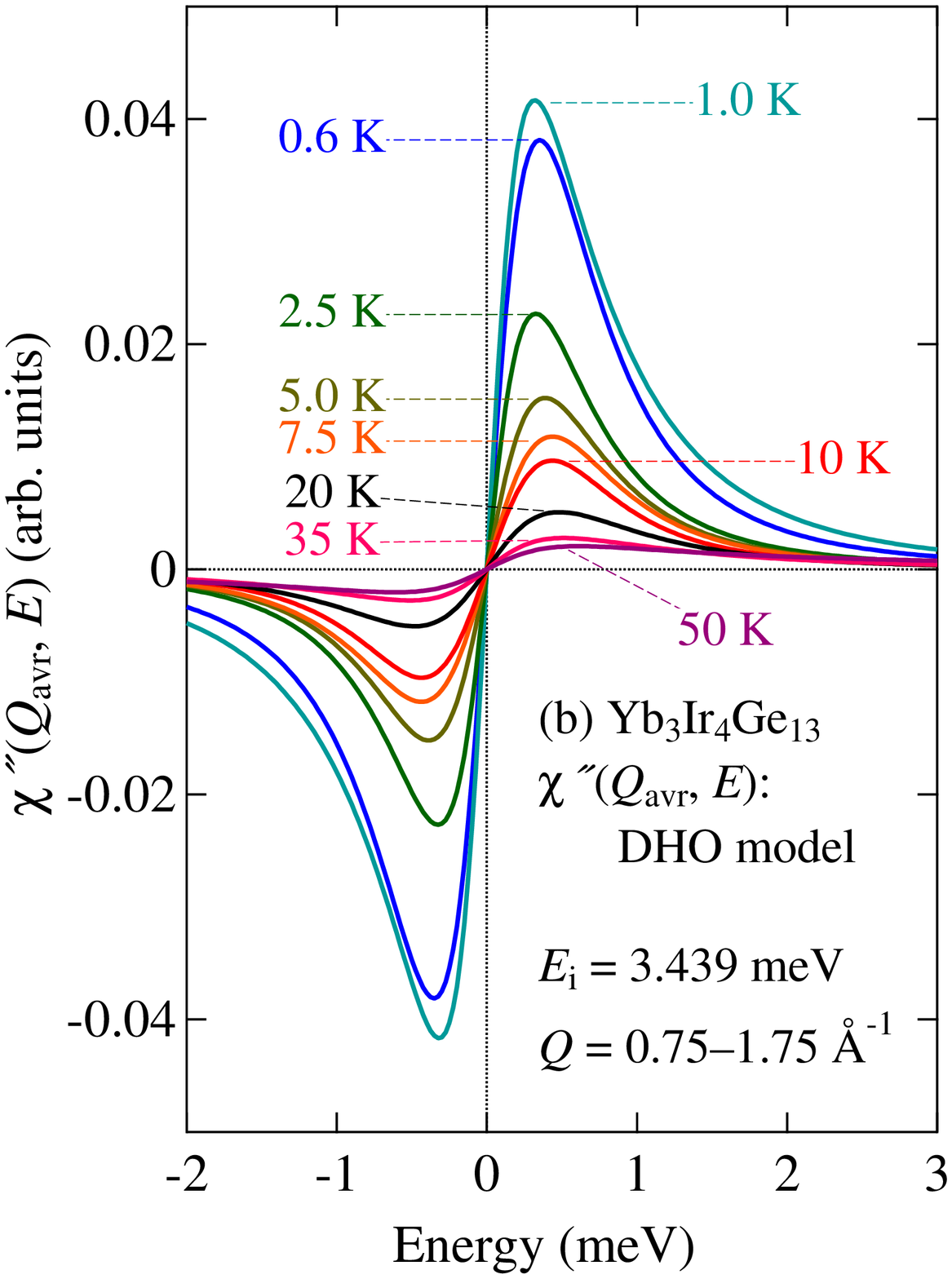}
\caption{\label{f_DHOparameters} (Color online) (a) Temperature dependence of $\chi'_0$, $E_0$, and $\gamma$ obtained in the fitting analysis based on the DHO model function $\chi''(Q_{\rm avr}, E)$.
(b) $\chi''(Q_{\rm avr}, E)$ calculated using the fitted parameters.}
\end{figure}

The INS signal up to $E=$ 3~meV is much broader than the instrumental resolution at the elastic scattering position. This fact allows us not to adopt a resolution convolution analysis. Thus, we simply fit the above function $S(Q_{\rm avr},E)$ to the data. Results of the least-squares fitting procedure are shown by black solid lines in Fig.~\ref{f_spectra}, and reproduce the measurement data satisfactorily. Temperature dependencies of the parameters $\chi'_0$, $E_0$, and $\gamma$, are summarized in Fig.~\ref{f_DHOparameters}
The parameter $\chi'_0$ corresponds to a $Q$-integrated static magnetic susceptibility, and increases with decreasing temperature below approximately 20~K. 
The parameter $E_0$ corresponding to a bare oscillator energy is 0.7~meV near the lowest measuring temperature. 
The spectral width parameter $\gamma$ is within 1.3~meV of the lowest measured temperature.
The fitting result to the 50-K data provides large uncertainties of the parameters because the magnitude of $\chi''(Q_{\rm avr}, E)$ is suppressed, broadened with respect to $E$, and multiplied by a large value of $n(E)+1$. This fact means that the magnetic scattering is no longer present above 50~K.

Figure~\ref{f_DHOparameters}(b) shows temperature dependence of $\chi''(Q_{\rm avr}, E)$ extracted from the fitting analysis. The $\chi''(Q_{\rm avr}, E)$ value is clearly enhanced below 20~K, not only at $E=0$ (corresponding to $\chi'_0$) but also in the whole $E$ range up to 3~meV. In addition, the magnitude $\chi''(Q_{\rm avr}, E)$ as well as $\chi'_0$ at 0.6~K are slightly smaller than those at 1~K within the present experimental accuracy. This fact is consistent with the maximum value of magnetic susceptibility at $T^\ast_{\rm mag} = 0.9$~K.

The $Q$ dependence of the magnetic excitation can be discussed on the basis of the magnetic cluster correlation. The $S(Q)$ of the low-energy excitation enhanced, shown in Fig.~\ref{f_IQE}(c) and \ref{f_IQE}(d), indicates the emergence of antiferromagnetic fluctuations. We simply analyzed the data using a scattering function form for a spin cluster~\cite{Spincluster_INS,RbMnMgF3_LAM80C,LaCoO3_Podlesnyak2008,LiGa0.95In0.05Cr4O8_Tanaka2018}, which is represented as 
\begin{eqnarray}
S(Q) = f^2(Q) \sum_j A_j \frac{\sin(QR_j)}{QR_j}
\label{eq_Cluster}
\end{eqnarray}
where $A_j$~($j \geq 1$) is a parameter of spin-spin correlation multiplied by is the coordination number of the spins located at the distance $R_j$ and $f(Q)$ is a magnetic form factor of the Yb$^{3+}$ ion~\cite{Pr_magff}. The distance between the nearest neighbors Yb ions is $R_1 = 4.45$~\AA, the 2nd neighbors $R_2 = 5.45$~\AA, the 3rd neighbors $R_3 = 8.3$~\AA, and the 4th neighbors $R_4 = 8.9$~\AA, which were estimated from the structural analysis of Lu$_3$Ir$_4$Ge$_{13}$~\cite{Oswald_Yb3Rh4Sn13-type}.
A least-squares fitting analysis based on the above $S(Q)$ form to the data at 0.6~K was performed, and the results are shown by the blue solid line in Fig.~\ref{f_IQE}(c). The determined parameters are 
 $A_0 = 0.0296$, 
 $A_1 = -0.028$,
 $A_2 = 0.030$,
 $A_3 =-0.022$, and
 $A_4 =0.011$.
The results for the 20~K data are also shown by a red solid line in Fig.~\ref{f_IQE}(d), the parameter for which are
 $A_0 =0.0170$,
 $A_1 =-0.012$,
 $A_2 =0.007$,
 $A_3 =-0.014$, and
 $A_4 =0.015$.
The negative value of $A_1$ indicates antiferromagnetic correlations between the nearest neighbor Yb ions. The decrease in the parameters with increasing temperature is consistent with the suppressions of the magnetic correlations. In the present analysis, we neglected the detailed crystal structure. For example, the nearest-neighbor Yb-ion distance is not a single value of $R_1 = 4.45$~\AA~assumed in the present model. Such non-uniform structure might cause a discrepancy between the observed $S(Q)$ and the model calculation, as seen in Fig.~\ref{f_IQE}(c) and \ref{f_IQE}(d). However, the overall feature is well reproduced by the present cluster model.

\section{\label{sec:level5}Discussion\protect\\}

Because the high-temperature magnetic susceptibility of Yb$_3$Ir$_4$Ge$_{13}$ is reproduced by the Curie--Weiss law, the Yb-ion valence is expected to be much closer to trivalent with $4f^{13}$ configuration, and divalent non-magnetic state is not dominant. For the monoclinic local symmetry ($C_2$) at the Yb site, four doublet crystal-electric-field (CEF) levels due to the $J=7/2$ state of the Yb$^{3+}$ ion are expected~\cite{CFHamiltonian_Walter,CFHamiltonian_Handbook_RareEarth}. There are three inequivalent Yb-ion sites in the unit cell, and thus we can expect several INS signals due to the excitations between the CEF levels. However, no clear CEF excitation peaks were detected in the present study on Yb$_3$Ir$_4$Ge$_{13}$, which is in contrast to the clear CEF peaks at approximately 7--9 and 28--39~meV of Ce$_3${\it T}$_4$Sn$_{13}$ ({\it T}: Co and Rh)~\cite{Ce3Co4Sn13_INS_Christianson,Ce3Rh4Sn13_INS_Adroja,Ce3Co4Sn13_INS_Iwasa,Ce3Rh4Sn13_INS_AMATERAS_ThALES}. The very broad magnetic scattering was observed in the $E$ range above the low-energy excitation of Yb$_3$Ir$_4$Ge$_{13}$. As demonstrated in the intensity integration over $E=$ 20--40~meV, this broad signal is also of magnetic-scattering origin. This fact indicates that the CEF levels are damped considerably owing to the hybridization with the conduction electrons, or the Yb-ion valence deviates slightly from trivalent at the INS measuring temperature. 

In contrast, the magnetic excitation below 3~meV was clearly observed. The ${\bm Q}$-integrated static magnetic susceptibility is closely related with the electrical resistivity because these are enhanced drastically below approximately 20~K. The magnetic moment of the Yb ions exhibits a short-range and short-life-time correlations, which is expected to scatter semimetal carriers in Yb$_3$Ir$_4$Ge$_{13}$. In other words, the local interactions from the carriers to the Yb-ion magnetic moments favor the paramagnetic state at temperature even much below the absolute value of the Weiss temperature ($-17$~K). In addition, the antiferromagnetic correlations were derived from the momentum dependence of the $S(Q)$ data [Figs.~\ref{f_IQE}(c) and \ref{f_IQE}(d)], which is consistent with the negative Weiss temperature. The characteristic energy of the magnetic fluctuations is $E_0=$ 0.7~meV and the inverse of life time $\gamma=1.3$~meV gives a maximum in ${\chi}''(Q_{\rm avr}, E)$ at 0.3~meV ($\approx$ 3~K). This characteristic energy of the antiferromagnetic fluctuations is very close to the specific-heat peak at $T^\ast_{\rm mag}$ = 0.9~K. This estimate is consistent with the thermodynamic signatures for the fragile magnetism~\cite{Yb3Ir4Ge13_fragilemagnetism}. We also noted above that the magnitude $\chi''(Q_{\rm avr}, E)$ and $\chi'_0$ at 0.7~K are slightly smaller than those at 1~K. This fact indicating a suppression of the magnetic fluctuation probably gives rise to the cusp-like anomaly in the magnetic susceptibility at $T^\ast_{\rm mag} = 0.9$~K. We can propose a significance of the antiferromagnetic fluctuation for origins of all the anomalies in the bulk properties.

We would like to compare the above-mentioned physical behaviors of Yb$_3$Ir$_4$Ge$_{13}$ to those of Yb$_3T_4$Ge$_{13}$ ($T$: Co and Rh)~\cite{Yb3RhCoIr4Ge13_PRB93}. The specific heat scaled by temperature of the Co/Rh-based compounds does not exhibit a peaking behavior in contrast to the peak at $T^\ast_{\rm mag}$ of the Ir-based compound. The magnetization of the Co/Rh-based compounds is very small, and the electrical resistivity is less temperature dependent than those of the Ir-based compound. These facts were explained by the valence fluctuation involving the nonmagnetic Yb$^{2+}$ state. The magnetic behaviors and low-temperature specific heat of Yb$_3$(Rh$_{1-x}$Ir$_x$)$_4$Ge$_{13}$ are enhanced for $x > 0.2$. The present magnetic excitation for Yb$_3$Ir$_4$Ge$_{13}$ is consistent with such an electronic evolution owing to the increase in Ir content. It is noteworthy that there are three inequivalent Yb-ion sites of Yb$_3$Ir$_4$Ge$_{13}$ in the $I4_1/amd$ space group, and the magnetic entropy at the temperature twice of $T^\ast_{\rm mag}$ is only $0.4 R \ln 2$~J/(mol-Yb~K)~\cite{Yb3Ir4Ge13_fragilemagnetism}. These facts may indicate that partial Yb ions become the magnetic trivalent state without forming a robust magnetic ordering. Such a nonuniform Yb-valence state is also in accordance with the less significant CEF excitation in the present INS data and the magnetically fluctuating state.

The low-temperature behaviors of Ce$_3${\it T}$_4$Sn$_{13}$ ({\it T}: Co, Rh, and Ir) are similar to those of Yb$_3$Ir$_4$Ge$_{13}$, although the crystal structures of these Sn-based compounds are slightly different from that of Yb$_3$Ir$_4$Ge$_{13}$. Ce$_3${\it T}$_4$Sn$_{13}$ ({\it T}: Co and Rh) are paramagnetic states even at 0.4--0.5~K, which is below the specific-heat peaks located near 1~K~\cite{Ce3Co4Sn13_Cornelius,Ce3Co4Sn13_LyleThomas,Ce3Co4Sn13_Slebarski_2012,Ce3Co4Sn13_Slebarski_2013,Ce3Rh4Sn13_Oduchi,Ce3Rh4Sn13_Kohler,Ce3Rh4Sn13_Gamza,Ce3Rh4Sn13_Xrayetc_Iwasa}, whereas a semiconductor Ce$_3$Ir$_4$Sn$_{13}$ were reported to undergo an antiferromagnetic phase transition at 0.6~K~\cite{Ce3Ir4Sn13_HF_Sato,Ce3Ir4Sn13_Nagoshi2005}. Electrical resistivity values of Ce$_3${\it T}$_4$Sn$_{13}$ ({\it T}: Co and Rh) are less dependent on temperature, and remain high even at low temperatures compared to the reference superconductor materials La$_3${\it T}$_4$Sn$_{13}$ ({\it T}: Co and Rh). It is noteworthy that the resistivity of Ce$_3${\it T}$_4$Sn$_{13}$ ({\it T}: Co and Rh) is enhanced below 15~K, and shows maxima at approximately 1~K. Recently, INS measurements were also performed for Ce$_3${\it T}$_4$Sn$_{13}$ ({\it T}: Co and Rh)~\cite{Ce3Co4Sn13_INS_Christianson,Ce3Rh4Sn13_INS_Adroja,Ce3Co4Sn13_INS_Iwasa,Ce3Rh4Sn13_INS_AMATERAS_ThALES}. These compounds exhibit clear CEF excitations at approximately 7--9 and 28--39~meV in their chiral $I2_13$ crystal-structure phase below 160 and 350~K of Ce$_3$Co$_4$Sn$_{13}$~\cite{R3Co4Sn13_BL8B} and Ce$_3$Rh$_4$Sn$_{13}$~\cite{Ce3Rh4Sn13_Xrayetc_Iwasa}, respectively. For these two compounds, low-energy magnetic excitation signals located below 1~meV were clearly observed below 15~K. These excitations were attributed to modification of a doublet CEF ground state of the Ce$^{3+}$ $4f^1$ configuration due to inter-site interaction or a hybridization with carriers. In the present INS study, we detected a similar magnetic response in Yb$_3$Ir$_4$Ge$_{13}$. Therefore, we propose that the strong magnetic fluctuations of these 3--4--13 systems play a role in the carrier scattering as seen in the resistivity increase below 15--20~K. As far as we know, it is novel that the magnetic fluctuation significantly influences the carrier transportation in semimetal materials. 
The magnetically ordering semiconductor Ce$_3$Ir$_4$Sn$_{13}$ is also a system to examine such a scenario of the relevance of magnetic fluctuations to the transport property. In addition, the resistivity data of Ce$_3${\it T}$_4$Sn$_{13}$ ({\it T}: Co and Rh) take the maximum values at approximately 1--2~K~\cite{Ce3Co4Sn13_rho_highP,Ce3Co4Sn13_Slebarski_2012,Ce3Rh4Sn13_Kohler}, while the saturation behavior without any significant maximum appears down to 0.1~K in the resistivity of  Yb$_3$Ir$_4$Ge$_{13}$~\cite{Yb3RhCoIr4Ge13_PRB93,Yb3Ir4Ge13_fragilemagnetism}. It is attractive to study on conduction-band structures in these 3--4--13 compounds, which cause the various carrier scattering phenomena associated with the characteristic magnetic fluctuations.

\section{\label{sec:level6}Conclusion\protect\\}

The magnetic excitation in Yb$_3$Ir$_4$Ge$_{13}$ has been revealed via INS experiments. The spectrum does not show any significant peaks corresponding to excitations between the CEF levels associated with the Yb $4f$ electronic state. In contrast, the ground state was found to provide clear magnetic excitation below 3~meV, which evolves below 20~K simultaneously with the reported steep upturn of the electrical resistivity with decreasing temperature~\cite{Yb3RhCoIr4Ge13_PRB93,Yb3Ir4Ge13_fragilemagnetism}. In addition, the antiferromagnetic correlation is seen in the momentum dependence. This low-energy magnetic excitation is also relevant to the specific-heat maximum at 0.9~K, and thus this result sheds light on these anomalies associated with the magnetic fluctuation in Yb$_3$Ir$_4$Ge$_{13}$.   

\begin{acknowledgments}
The experiment at BL14, the Materials and Life Science Facility, J-PARC, was based on approved proposals (No.~2018A0150).
M.~Kofu is acknowledged for conducting the measurement at BL14. The J-PARC technical staffs are also acknowledged for assistance in the measurement using the $^3$He refrigerator at BL02.
K. I. was supported in part by the Japan Society for the Promotion of Science, KAKENHI Grant No.~JP17H05209 [Scientific Research on Innovative Areas ``3D Active-Site Science"] and JP18H01182 [Scientific Research (B)].
C.-L.~H., B. K. R. and E. M. acknowledge support from the Gordon and Betty Moore Foundation EPiQS Initiative through grant GBMF 4417.
\end{acknowledgments}


\begin{thebibliography}{9}

\bibitem{denseKondoEffect}
T. Kasuya and T. Saso (Eds.),
{\it Theory of Heavy Fermions and Valence Fluctuations}, (Springer, Heidelberg, 1985).
\bibitem{HF_Grewe_Steglich} N. Grewe and F. Steglich, 
in {\it Handbook on the Physics and Chemistry of Rare Earths}, eds. K. A. Gschneidner, Jr. and L. Eyring (North-Holland, Amsterdam, 1991), Vol. 14, pp. 343-474.







\bibitem{Oswald_Yb3Rh4Sn13-type} I. W. H. Oswald, B. K. Rai, G. T. McCandless, E. Morosan, and J. Y. Chan,
Cryst. Eng. Comm $\bf{19}$, 3381 (2017).

\bibitem{Ce3Co4Sn13_Cornelius} A. L. Cornelius, A. D. Christianson, J. L. Lawrence, V. Fritsch, E. D. Bauer, J. L. Sarrao, J. D. Thompson, and P. G. Pagliuso,
Physics B $\bf{378}$--$\bf{380}$, 113 (2006).
\bibitem{Ce3Co4Sn13_LyleThomas} E. L. Thomas, H.-O. Lee, A. N. Bankston, S. MaQuilon, P. Klavins, M. Moldovan, D. P. Young, Z. Fisk, and J. Y. Chan, 
J. Solid State Chem. $\bf{179}$, 1642 (2006).
\bibitem{Ce3Co4Sn13_Slebarski_2012} A. {\'S}lebarski, B. D. White, M. Fija{\l}kowski, J. Goraus, J. J. Hamlin, and M. B. Maple,
Phys. Rev. B $\bf{86}$, 205113 (2012).
\bibitem{Ce3Co4Sn13_Slebarski_2013} A. {\'S}lebarski and J. Goraus,
Phys. Rev.  B $\bf{88}$, 155122 (2013).

\bibitem{Yb3RhCoIr4Ge13_PRB93} B. K. Rai, I. W. H. Oswald, J. Y. Chan, and E. Morosan,
Phys. Rev. B $\bm{93}$, 035101 (2016).
\bibitem{Yb3Ir4Ge13_fragilemagnetism} B. K. Rai, I. W. H. Oswald, W. Ban, C.-L. Huang, V. Loganathan, A. M. Hallas, M. N. Wilson, G. M. Luke, L. Harriger, Q. Huang, Y. Li, S. Dzsaber, J. Y. Chan, N. L. Wang, S. Paschen, J. W. Lynn, A. H. Nevidomskyy, P. Dai, Q. Si, and E. Morosan,
accepted by Phys. Rev. B.
\bibitem{RE3Ir4Ge13_synthesis} B. K. Rai, I. W. H. Oswald, J. K. Wang, G. T. Mc-Candless, J. Y. Chan, and E. Morosan, Chem. Matter. $\bm{27}$, 2494 (2015).

\bibitem{AMATERAS_JPSJ} K. Nakajima, S. Ohira-Kawamura, T. Kikuchi, M. Nakamura, R. Kajimoto, Y. Inamura, N. Takahashi, K. Aizawa, K. Suzuya, K. Shibata, T. Nakatani, K. Soyama, R. Maruyama, H. Tanaka, W. Kambara, T. Iwahashi, Y. Itoh, T. Osakabe, S. Wakimoto, K. Kakurai, F. Maekawa, M. Harada, K. Oikawa, R. E. Lechner, F. Mezei, and M. Arai,
J. Phys. Soc. Jpn. ${\bf 80}$, SB028  (2011).
\bibitem{UTSUSEMI} Y. Inamura, T. Nakatani, J. Suzuki, and T. Otomo,
J. Phys. Soc. Jpn. ${\bf 82}$, SA031 (2013).

\bibitem{Pr_magff} E. J. Lisher and J. B. Forsyth,
Acta. Cryst. A \textbf{27}, 545 (1971).

\bibitem{Spincluster_INS}
A. Furrer and H. U. G\"udel,
Phys. Rev. Lett. \textbf{39}, 657 (1977) and J. Magn. Magn. Mater. \textbf{14}, 256 (1979).
\bibitem{RbMnMgF3_LAM80C}
K. Iwasa, K. H. Andersen, M. Takahashi, and H. Ikeda,
J. Phys. Soc. Jpn. \textbf{63}, 2862 (1994).
\bibitem{LaCoO3_Podlesnyak2008}
A. Podlesnyak, M. Russina, A. Furrer,   A. Alfonsov, E.  Vavilova, V.  Kataev, B. Bu\"chner, Th.  Stra\"ssle,  E. Pomjakushina, K.  Conder, and D. I.  Khomskii,
Phys. Rev. Lett.  \textbf{101}, 247603 (2008).
\bibitem{LiGa0.95In0.05Cr4O8_Tanaka2018}
Y. Tanaka,  R.  Wawrzy\'nczak, Manh Duc  Le, T.  Guidi,  Y.  Okamoto, T.  Yajima, Z.  Hiroi, M.  Takigawa, and G. J. Nilsen,
J. Phys. Soc. Jpn. \textbf{87}, 073710 (2018).

\bibitem{CFHamiltonian_Walter} U. Walter,
J. Phys. Chem. Solids $\bm{45}$, 401 (1984).
\bibitem{CFHamiltonian_Handbook_RareEarth} C. G\"oller-Walrand and K. Binnemans,
in {\it Handbook on the Physics and Chemistry of Rare Earths}, edited by K. A. Gschneidner, Jr. and L. Eyring (Elsevier Science B. V., North-Holland, 1996) Vol.~23, p.~121.

\bibitem{Ce3Co4Sn13_INS_Christianson} A. D. Christianson, J. S. Gardner, H. J. Kang, J.-H. Chung, S. Bobev, J. L. Sarrao, and J. M. Lawrence,
J. Magn. Magn. Mater. $\bm{310}$, 266 (2007).
\bibitem{Ce3Co4Sn13_INS_Iwasa} K. Iwasa, Y. Otomo, K. Suyama, K. Tomiyasu, S. Ohira-Kawamura, K. Nakajima, and J.-M. Mignot,
Phys. Rev. B $\bf{95}$, 195156 (2017).


\bibitem{Ce3Rh4Sn13_INS_Adroja} D. T. Adroja, A. M. Strydom, A. P. Murani, W. A. Kockelmann, and A. Fraile,
Physica B $\bm{403}$, 898 (2008).


\bibitem{Ce3Rh4Sn13_INS_AMATERAS_ThALES}
K. Iwasa, K. Suyama, Y. Otomo, K. Tomiyasu, S. Raymond, P. Steffens, and J.-M. Mignot (unpublished).








\bibitem{Ce3Rh4Sn13_Oduchi} Y. \=Oduchi, C. Tonohiro, A. Thamizhavel, H. Nakashima, S. Morimoto, T. D. Matsuda, Y. Haga, K. Sugiyama, T. Takeuchi, R. Settai, M. Hagiwara, and Y. \=Onuki,
J. Magn. Magn. Mater.  $\bf{310}$, 249 (2007).
\bibitem{Ce3Rh4Sn13_Kohler} U. K\"ohler, A. P. Pikul, N. Oeschler, T. Westerkamp, A. M. Strydom and F. Steglich, 
J. Phys.: Condens. Matter $\bf{19}$, 386207 (2007).
\bibitem{Ce3Rh4Sn13_Gamza} M. Gam{\. z}a, W. Schnelle, A. {\'S}lebarski, U. Burkhardt, R. Gumeniuk, and H. Rosner, 
J. Phys.: Condens. Matter $\bf{20}$, 395208 (2008).
\bibitem{Ce3Rh4Sn13_Xrayetc_Iwasa} K. Suyama, K. Iwasa, Y. Otomo, K. Tomiyasu, H. Sagayama, R. Sagayama, H. Nakao, R. Kumai, Y. Kitajima, F. Damay, J.-M. Mignot, A. Yamada, T. D. Matsuda, and Y. Aoki,
Phys. Rev. B. $\bf{97}$, 235138 (2018).

\bibitem{Ce3Ir4Sn13_HF_Sato} H. Sato, T. Fukuhara, S. Iwakawa, Y. Aoki, I. Sakamoto, S. Takayanagi, and N. Wada,
Physica B $\bf{186-188}$, 630 (1993).
\bibitem{Ce3Ir4Sn13_Nagoshi2005} C. Nagoshi, H.  Sugawara, Y.  Aoki, S.  Sakai, M.  Kohgi, H. Sato, T.  Onimaru, and T. Sakakibara,
Physica B $\bf{359-361}$, 248 (2005).


\bibitem{R3Co4Sn13_BL8B} Y. Otomo, K. Iwasa, K. Suyama, K. Tomiyasu, H. Sagayama, R. Sagayama, H. Nakao, R. Kumai, and Y. Murakami,
Phys. Rev. B. $\bm{94}$, 075109 (2016).

\bibitem{Ce3Co4Sn13_rho_highP} J. R. Collave, H. A. Borges, S. M. Ramos, E. N. Hering, M. B. Fontes, E. Baggio-Saitovitch, L. Mendon\c{c}a-Ferreira, E. M. Bittar, and P. G. Pagliuso,
J. Appl. Phys. $\bm{117}$, 17E307 (2015).





\end{thebibliography}

\end{document}